\documentclass[aps,prl,twocolumn,superscriptaddress,showpacs]{revtex4}

\usepackage{graphics}
\usepackage{amsmath}
\usepackage{bm}

\begin{document}

\title{Asymmetric Zero-Bias Anomaly for Strongly Interacting Electrons in
  One Dimension}

\author{K. A. Matveev}

\affiliation{Materials Science Division, Argonne National Laboratory, 
Argonne, Illinois 60439, USA}

\author{A. Furusaki}

\affiliation{Condensed Matter Theory Laboratory, RIKEN, Wako, Saitama
  351-0198, Japan} 

\author{L. I. Glazman}

\affiliation{Theoretical Physics Institute, University of Minnesota, 
Minneapolis, Minnesota 55455, USA}

\date{\today}

\begin{abstract}
  We study a system of one-dimensional electrons in the regime of strong
  repulsive interactions, where the spin exchange coupling $J$ is small
  compared with the Fermi energy, and the conventional Tomonaga-Luttinger
  theory does not apply.  We show that the tunneling density of states has
  a form of an asymmetric peak centered near the Fermi level.  In the
  spin-incoherent regime, where the temperature is large compared to $J$,
  the density of states falls off as a power law of energy $\varepsilon$
  measured from the Fermi level, with the prefactor at positive energies
  being twice as large as that at the negative ones.  In contrast, at
  temperatures below $J$ the density of states forms a split peak with
  most of the weight shifted to negative $\varepsilon$.
\end{abstract}


\maketitle

The discovery of conductance quantization in quantum wires \cite{old} has
stimulated interest in transport properties of one-dimensional conductors.
From theoretical point of view, these systems are interesting because in
one dimension interacting electrons form the so-called Luttinger liquid
\cite{giamarchi}, with properties very different from the conventional
Fermi liquids.  A number of non-trivial properties of the Luttinger
liquid, such as the power-law dependence of the tunneling density of
states on energy and temperature, have been recently observed
experimentally \cite{yacoby1,bockrath,yao}.

It is important to note that the Luttinger-liquid picture \cite{giamarchi}
describes only the low-energy properties of the system.  Quantitatively,
this means that all the important energy scales, such as the temperature
$T$, must be much smaller than the typical bandwidth of the problem.  In a
system of spin-$\frac12$ electrons the charge and spin excitations
propagate at different velocities \cite{dzyaloshinskii}, resulting in two
bandwidth parameters.  In the non-interacting case both bandwidths are
equal to the Fermi energy $E_F=(\pi\hbar n)^2/8m$, where $n$ is the
electron density and $m$ is the effective mass.  Repulsive interactions
between electrons increase the charge bandwidth $D_\rho$ and decrease the
spin bandwidth $D_\sigma$.  At moderate interaction strength both
bandwidths remain of the order of Fermi energy $E_F$, and the
Luttinger-liquid picture applies at $T\ll E_F$.  On the other hand, if the
interactions are strong, the exchange coupling of electron spins $J$ is
strongly suppressed, and $D_\sigma\sim J \ll D_\rho$.  As a result the
Luttinger-liquid picture applies only at $T\ll J$, and there appears an
interesting new regime when the temperature is in the range $J\ll T\ll
D_\rho$.  In this regime the temperature does not strongly affect the
charge excitations in the system, but completely destroys the ordering of
the electron spins.

A number of interesting phenomena have been predicted in this so-called
\emph{spin-incoherent regime.}  The destruction of spin order may be
responsible \cite{matveev} for the anomalous quantization of conductance
in the experiments \cite{reilly-thomas}.  Furthermore, contrary to the
conventional Luttinger-liquid picture, the tunneling density of states may
show a power-law \emph{peak\/} near the Fermi level \cite{cheianov,fiete}
even in the case of repulsive interactions,
\begin{equation}
  \label{eq:cheianov-fiete_result}
  \nu(\varepsilon) \propto\frac{|\varepsilon|^{\frac{1}{4K_\rho}-1}}
  {\sqrt{\ln(D_\rho/|\varepsilon|)}},
  \quad J\ll T\ll|\varepsilon|\ll D_\rho.
\end{equation}
Here $\varepsilon$ is the energy of the tunneling electron measured from
the Fermi level.  The result (\ref{eq:cheianov-fiete_result}) was first
obtained \cite{cheianov} for the Hubbard model in the limit of strong
on-site repulsion $U\to\infty$.  In this case the interactions have a very
short range, resulting in $K_\rho=1/2$ and $\nu\propto
[|\varepsilon|\ln(D_\rho/|\varepsilon|)]^{-1/2}$.  For longer-range
interactions $K_\rho$ is below 1/2, but as long as $K_\rho>1/4$, the
density of states has a peak at low energies.

A similar enhancement of the density of states at $\varepsilon\ll D_\rho$ was
predicted earlier \cite{penc1} in the strongly interacting limit of the
Hubbard model at zero temperature,
\begin{equation}
  \label{eq:penc_result}
   \nu(\varepsilon) \propto |\varepsilon|^{-3/8},
  \qquad
  T=0,\,
  J\ll |\varepsilon|\ll D_\rho.
\end{equation}
Here the effective exchange coupling of electron spins $J\sim t^2/U$, the
bandwidth $D_\rho\sim t$, and $t$ is the hopping matrix element in the
Hubbard model.  The different power-law behaviors of the density of states
(\ref{eq:cheianov-fiete_result}) and (\ref{eq:penc_result}) point to the
non-trivial physics developing when the temperature $T$ is lowered below
the exchange $J$, even if they are both small compared to the energy
$\varepsilon$.

In this paper we develop a unifying theory, which enables one, in
principle, to obtain the density of states in a system of strongly
interacting one-dimensional electrons at arbitrary ratio $T/J$, as long as
$T,J\ll\varepsilon$.  In the spin-incoherent case, $T\gg J$, our theory
reproduces the result (\ref{eq:cheianov-fiete_result}).  Furthermore, we
show that true asymptotic behavior of $\nu(\varepsilon)$ at low energies
is given by Eq.~(\ref{eq:cheianov-fiete_result}) even at $T=0$.  Most
importantly, in both cases the peak in $\nu(\varepsilon)$ is very
asymmetric.  In particular, the 3/8-power law (\ref{eq:penc_result})
appears for short-range interactions at moderately low positive energies
$\varepsilon$, but never at $\varepsilon<0$.

Our approach is based on the fact that at strong interactions, when
$J/D_\rho\to0$, the Hamiltonian of the system of one-dimensional electrons
separates into a sum of two contributions, $H=H_\rho+H_\sigma$, describing
the charge and spin degrees of freedom, respectively.  This result was
first obtained by Ogata and Shiba \cite{ogata} in the $U/t\to\infty$ limit
of the Hubbard model.  In the case of quantum wires at low electron
densities a similar separation of charge and spin degrees of freedom was
discussed in Ref.~\onlinecite{matveev}.  This separation occurs whenever
the repulsive interactions are so strong that even electrons with opposite
spins do not occupy the same point in space.  In this case the electrons
are well separated from each other, and their spins form a spin chain.
The residual coupling $J$ of the neighboring spins is weak, $J\ll E_F$,
and the spin excitations are described by the isotropic Heisenberg model
\begin{equation}
  \label{eq:spin-chain}
  H_\sigma = \sum_l J\, {\bm S}_{l}\cdot{\bm S}_{l+1}.
\end{equation}
Since the Pauli principle is effectively enforced by the interactions even
for electrons with opposite spins, the charge part $H_\rho$ of the
Hamiltonian can be written in terms of spinless fermions (holons).  In the
case of the Hubbard model \cite{ogata} the holons are non-interacting,
because two holons never occupy the same lattice site, and the interaction
range does not extend beyond single site.  On the other hand, if the
original interaction between electrons has non-zero range, the holons do
interact.  Since we are only interested in the density of states at
energies $\varepsilon$ low compared to the bandwidth $D_\rho$, the holon
Hamiltonian can be bosonized,
\begin{eqnarray}
  \label{eq:H_rho}
  H_\rho=\int\frac{\hbar u_\rho}{2\pi}\left[K (\partial_x\theta)^2 
                         +K^{-1}(\partial_x\phi)^2\right]dx.
\end{eqnarray}
Here $u_\rho$ is the speed of charge excitations and $K$ is related to the
standard Luttinger-liquid parameter $K_\rho$ for the charge modes in an
interacting electron system \cite{giamarchi} as $K=2K_\rho$, with the
factor of 2 originating from the different definition of the bosonic
fields $\phi$ and $\theta$.

To find the tunneling density of states, one needs an expression for the
electron creation and annihilation operators.  The electron annihilation
operator $\psi_\sigma(x)$ affects both the charge and spin degrees of
freedom: it destroys a holon at point $x$ and removes a site with spin
$\sigma$ from the spin chain (\ref{eq:spin-chain}).  Building on the ideas
of Refs.~\onlinecite{matveev}, \onlinecite{fiete}, \onlinecite{penc1}, and
\onlinecite{brazovskii} we write $\psi_\sigma(x)$ as
\begin{equation}
  \label{eq:annihilation}
  \psi_\sigma(x)=\frac{e^{\pm i[k_Fx+\phi(x)]-i\theta(x)}}
                      {(2\pi\alpha)^{1/2}}\,
                    Z_{l,\sigma}
                 \Big|_{l=\frac{k_F x+\phi(x)}{\pi}}.
\end{equation}
The first factor is the bosonized form of an operator destroying a holon
on the right- or left-moving branch, depending on the sign in the
exponent.  [Here $\alpha=\hbar u_\rho/D_\rho$ is the short-distance
cutoff; the holon Fermi momentum is related to the mean electron density
$n$ as $k_F=\pi n$.]  The operator $Z_{l,\sigma}$ introduced in
Ref.~\onlinecite{penc1}, removes site $l$ with spin $\sigma$ from the spin
chain.  It is important to note that Eq.~(\ref{eq:annihilation}) properly
accounts for the fact \cite{matveev,fiete} that charge modes shift the
spin chain at point $x$ by $\delta l=\phi(x)/\pi$ with respect to its
average position $\bar l=k_Fx/\pi$.

It is convenient to express the operators $Z_{l,\sigma}$ in terms of their
Fourier transforms $z_\sigma(q)$, where the momentum $q$ is defined on a
lattice and assumes values between $-\pi$ and $\pi$.  Then for the
operator destroying a right-moving holon we obtain
\begin{equation}
  \label{eq:annihilation_Fourier}
  \psi_\sigma^R(x)=\int_{-\pi}^\pi \frac{dq}{2\pi}\, z_\sigma(q)
                  \frac{e^{ik_F(1+\frac{q}{\pi})x}}{(2\pi\alpha)^{1/2}}\,
                  e^{i(1+\frac{q}{\pi})\phi(x)-i\theta(x)}.
\end{equation}
In addition to the holon destruction operator $e^{i(\phi-\theta)}$ the
integrand contains a factor $e^{i(q/\pi)\phi}$.  Similar factors
appear when bosonization is applied to the problem of X-ray-edge
singularity \cite{schotte,fiete2}, where they represent the effect of
the core-hole potential on the electronic wavefunctions.  More
specifically, a core hole with the scattering phase shift $\delta$
adds a factor $e^{i(2\delta/\pi)\phi}$ to the fermion operator.  Thus
according to Eq.~(\ref{eq:annihilation_Fourier}) the electron
tunneling process that changes the momentum of the spin chain by $q$
also adds a scattering phase shift for the holons $\delta=q/2$.  This
observation is consistent with the fact \cite{penc1} that for the
state of momentum $Q$ of the spin chain, the periodic boundary
conditions for the holons acquire a phase factor $e^{iQ}$.

The tunneling density of states $\nu(\varepsilon)$ can be computed as
imaginary part of the electron Green's function.  We first consider the
limit of zero temperature.  At $|\varepsilon|\gg J$ one can neglect the
time dependence of the operators $z_\sigma(q)$ and obtain
\begin{equation}
  \label{eq:nu_zeroT}
  \nu_\sigma^\pm(\varepsilon) = \nu_0
      \int_{-\pi}^\pi \frac{dq}{2\pi}\, 
      \frac{c_\sigma^\pm(q)}{\Gamma\big(\lambda(q)+1\big)}
      \left(\frac{|\varepsilon|}{D_\rho}\right)^{\lambda(q)}.
\end{equation}
for positive and negative $\varepsilon$.  Here $\nu_0=(\pi\hbar u_\rho)^{-1}$
and the exponent $\lambda(q)$ is given by
\begin{equation}
  \label{eq:lambda}
  \lambda(q) = \frac12\left[\left(1+\frac{q}{\pi}\right)^2 K 
                             + \frac{1}{K}\right] - 1.
\end{equation}
We have also defined
\begin{subequations}
  \label{eq:c}
\begin{eqnarray}
  c_\sigma^+(q) &=& 
      \sum_l\langle Z_{l,\sigma}^{} Z_{0,\sigma}^\dagger\rangle\, e^{-iql},
\\
  c_\sigma^-(q) &=& 
      \sum_l\langle Z_{0,\sigma}^\dagger Z_{l,\sigma}^{}\rangle\, e^{-iql},
\end{eqnarray}
\end{subequations}
with averaging performed over the ground state of the spin chain
(\ref{eq:spin-chain}).  

At $|\varepsilon|\ll D_\rho$ the dominant contribution to the integral in
Eq.~(\ref{eq:nu_zeroT}) comes from the vicinity of its lower limit,
\mbox{$q=-\pi$}.  The $|\varepsilon|/D_\rho\to0$ asymptote has the form
\begin{equation}
  \label{eq:nu_zeroT_asymptotics}
  \nu_\sigma^\pm(\varepsilon) = \nu_0
       \sqrt{\frac{\pi}{8 K}}
       \frac{c_\sigma^\pm(\pi)}{\Gamma\left(\frac{1}{2K}\right)}
       \left(\frac{|\varepsilon|}{D_\rho}\right)^{\frac{1}{2K}-1}
       \frac{1}{\sqrt{\ln\frac{D_\rho}{|\varepsilon|}}}.
\end{equation}
This result assumes that the functions (\ref{eq:c}) do not vanish at
$q=\pm\pi$.

It is important to point out that even though we so far assumed $T=0$, the
low-energy asymptote (\ref{eq:nu_zeroT_asymptotics}) agrees with
Eq.~(\ref{eq:cheianov-fiete_result}), rather than
Eq.~(\ref{eq:penc_result}).  Indeed, in the case of the Hubbard model the
holons are non-interacting, the parameter $K=1$, and the density of states
behaves as $\nu^\pm(\varepsilon)\propto 1/\sqrt{\varepsilon}$, instead of
the 3/8-power law (\ref{eq:penc_result}).

\begin{figure}[t]
 \resizebox{.46\textwidth}{!}{\includegraphics{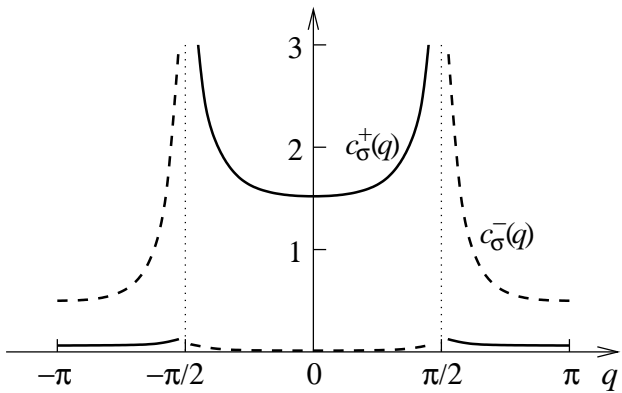}}
\caption{\label{fig:c} Sketch of the dependences $c^+_\sigma(q)$ and
  $c^-_\sigma(q)$ at zero temperature (solid and dashed lines,
  respectively), based on the numerical results of
  Ref.~\onlinecite{penc1}.}
\end{figure}

To resolve this disagreement, we first notice that at $K=1$ our
Eqs.~(\ref{eq:nu_zeroT}) and (\ref{eq:lambda}) are essentially equivalent
to Eqs.~(4), (10), and (13) of Ref.~\onlinecite{penc1}.  In the notations
of Ref.~\onlinecite{penc1} the functions (\ref{eq:c}) are given by
$c^+_\sigma(q)=(N+1)C_{\sigma,N}(\pi-q)$ and
$c^-_\sigma(q)=(N-1)D_{\sigma,N}(\pi-q)$ in the limit when the number of
sites $N$ in the spin chain is infinite.  These functions have been
computed numerically \cite{penc1}, and are shown schematically in
Fig.~\ref{fig:c}.  It is worth noting that the function $c^+_\sigma(q)$ is
numerically small at $\pi/2<|q|<\pi$, with $c^+_\sigma(\pm \pi)\approx
0.045$, whereas $c_\sigma^-(\pm\pi)\approx 0.46$ \cite{private}.  This
indicates that at very low energies $\varepsilon$ (but still
$\varepsilon\gg J$) the peak (\ref{eq:nu_zeroT_asymptotics}) of the
density of states is very asymmetric, with the tail below the Fermi level
being higher than the one above it by an order of magnitude.

Given the numerical smallness of the leading asymptote
(\ref{eq:nu_zeroT_asymptotics}) of the density of states at positive
$\varepsilon$, it is worth considering the subleading contributions to the
integral in Eq.~(\ref{eq:nu_zeroT}).  They come from the values of $q$ in
the range $-\pi/2<q<\pi/2$, where $c^+_\sigma(q)$ is of order one.  Taking
into account the fact that $c_\sigma^+(q)$ diverges as
$\chi/\sqrt{q+\pi/2}$ with $\chi\approx 0.8$ at $q\to-\pi/2+0$
\cite{penc1,sorella,unpublished}, we find
\begin{equation}
  \label{eq:nu_zeroT_subleading}
  \tilde\nu^+_\sigma(\varepsilon) = \nu_0
       \frac{\chi}{\sqrt{2 K}\,
                   \Gamma\left(\frac{1}{2K}+\frac{K}{8}\right)}
       \!
       \left(\frac{\varepsilon}{D_\rho}\right)^{\frac{1}{2K}+\frac{K}{8}-1}
       \frac{1}{\sqrt{\ln\frac{D_\rho}{\varepsilon}}}.
\end{equation}
At $K=1$ the exponent in Eq.~(\ref{eq:nu_zeroT_subleading}) becomes
$-3/8$, in agreement with Eq.~(\ref{eq:penc_result}).

At low positive energies the density of states can be treated as the sum
of the contributions $\nu^+_\sigma$ and $\tilde\nu^+_\sigma$, given by
Eqs.~(\ref{eq:nu_zeroT_asymptotics}) and (\ref{eq:nu_zeroT_subleading}).
The subleading contribution $\tilde\nu^+_\sigma$ diverges less rapidly
than $\nu^+_\sigma$ at $\varepsilon/D_\rho\to0$, but with the numerical
coefficient that is larger by a factor of about 20.  Thus for practical
purposes the peak in the density of states $\nu_\sigma(\varepsilon)$ is
given by $\tilde\nu_\sigma^+(\varepsilon)$,
Eq.~(\ref{eq:nu_zeroT_subleading}), at positive $\varepsilon$, and by
$\nu_\sigma^-(\varepsilon)$, Eq.~(\ref{eq:nu_zeroT_asymptotics}), at
negative $\varepsilon$.

We now turn to the spin-incoherent regime, $T\gg J$.  Assuming that
$|\varepsilon|\gg T$, one can still use Eq.~(\ref{eq:nu_zeroT}), however,
the definitions (\ref{eq:c}) of the functions $c^\pm_\sigma(q)$ should now
assume ensemble averaging.  At $T\gg J$ the functions $c^\pm_\sigma(q)$
can be easily computed by using the following simple argument
\cite{serhan}.  Given that the operators $Z_{l,\sigma}^{}$ and
$Z^\dagger_{l,\sigma}$ remove and add a site with spin $\sigma$ at
position $l$, it is clear that the ensemble average $\langle
Z_{0,\sigma}^\dagger Z_{l,\sigma}^{}\rangle$ equals the probability of all
the spins on sites $0,1,\ldots,l$ being $\sigma$.  At $J\ll T$ the spins
are completely random, so $\langle Z_{0,\sigma}^\dagger
Z_{l,\sigma}^{}\rangle=1/2^{|l|+1}$.  Similarly, $\langle Z_{l,\sigma}^{}
Z_{0,\sigma}^{\dagger}\rangle=1/2^{|l|}$.  Then the definitions
(\ref{eq:c}) give
\begin{equation}
  \label{eq:c_incoherent}
  c^+_\sigma(q)=2c^-_\sigma(q)=\frac{3}{5-4\cos q}.
\end{equation}
The expression for $c^-_\sigma(q)$ is equivalent to the result for
$D_{\sigma,N}(Q)$ found in Ref.~\onlinecite{serhan}.

It is important to point out that $c^+_\sigma(q)$ and $c^-_\sigma(q)$
differ by a factor of 2. As a result, the density of states
(\ref{eq:nu_zeroT_asymptotics}) has a clear asymmetry:
$\nu_\sigma(\varepsilon)=2\nu_\sigma(-\varepsilon)$ at $T\ll \varepsilon
\ll D_\rho$.  The physical meaning of this result is very simple: the
probabilities of adding an electron with spin $\sigma$ at energy
$\varepsilon$ and removing one at $-\varepsilon$ differ by a factor of 2
because the electron that is being removed has the correct spin with
probability 1/2.

The tunneling density of states can be studied experimentally by measuring
the $I$-$V$ characteristics of tunneling junctions in which one of the
leads is a quantum wire.  When the electron density in the wire is
sufficiently low, the exchange coupling is expected to be exponentially
weak, and the regime $J\sim T\ll D_\rho$ can be achieved \cite{matveev}.
In the experiment \cite{yacoby2} such a measurement was performed in a
situation where the second lead is another quantum wire.  By applying
magnetic field the authors have been able to observe momentum-resolved
tunneling.  To measure the density of states, it is more convenient to
make a point junction from a metal tip to the side of the quantum wire.
Contrary to the expectations based on the Luttinger-liquid theory, we
predict that the $I$-$V$ characteristic of such a junction will be very
asymmetric with respect to reversal of the bias when $J\ll eV\ll D_\rho$.
The dependence $\nu_\sigma(eV)\propto dI/dV$ should have a peak at low
bias at $K_\rho>1/4$, or a dip at $K_\rho<1/4$, but the asymmetry should
be observed in either case.  This asymmetry will indicate that
electron-electron scattering processes have become strong enough to
suppress the exchange coupling $J$ of electron spins.

The predicted asymmetry of the density of states is caused by the
non-trivial interplay of the spin and charge degrees of freedom.  As a
result, the asymmetry should disappear in a polarizing magnetic field,
$\mu_B B\gg T,J$.  Indeed, if all spins are fixed in the $\uparrow$
direction, one easily finds $c^\pm_\uparrow(q)=2\pi\delta(q)$.  Then
according to Eq.~(\ref{eq:nu_zeroT}) the density of states
$\nu_\uparrow(\varepsilon)\sim\nu_0 (|\varepsilon|/D_\rho)^{\lambda(0)}$.
This result recovers the standard Luttinger liquid suppression of the
density of states \cite{giamarchi} and shows no asymmetry around the Fermi
level.

\begin{figure}[t]
 \resizebox{.46\textwidth}{!}{\includegraphics{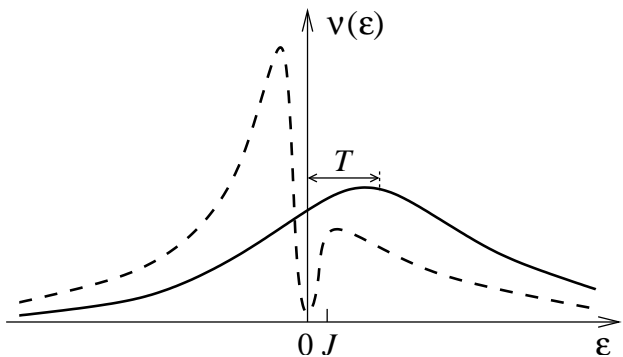}}
\caption{\label{fig:nu} Sketch of the tunneling density of states
  $\nu(\varepsilon)$ in two regimes: high temperature $T\gg J$ (solid
  line), and low temperature $T\ll J$ (dashed line).  At $T\gg J$ and
  $\varepsilon\gg T$ we predict $\nu(\varepsilon)=2\nu(-\varepsilon)$.  As
  the temperature is lowered below the exchange constant $J$, the density
  of states at $\varepsilon<0$ grows by about a factor of 3.  Conversely,
  at $\varepsilon>0$ the density of states decreases dramatically.  At
  $|\varepsilon|\ll J$ the standard Luttinger liquid effects give rise to
  power-law suppression of the density of states \cite{giamarchi}, leading
  to a dip at low $\varepsilon$.}
\end{figure}

Conductance of quantum wires is expected to depend strongly on the ratio
$J/T$ \cite{matveev}.  Our theory provides a new way to probe this ratio
by observing the asymmetry of the density of states $\nu(\varepsilon)$.
The signature of the spin-incoherent regime, $J/T\ll 1$, is the doubling
of the density of states at positive energies, compared to the negative
ones, $\nu(\varepsilon)=2\nu(-\varepsilon)$.  At $J/T\gg1$ the asymmetry
is inverted, $\nu(-\varepsilon)>\nu(\varepsilon)$.  This evolution of the
density of states is described by Eq.~(\ref{eq:nu_zeroT}), where the
temperature dependence is contained in the functions $c^\pm_\sigma(q)$.
Using Eq.~(\ref{eq:c}) one can easily check that in the absence of
magnetic field the integral of $c^+_\sigma(q)$ over all $q$ is always
larger than the integral of $c^-_\sigma(q)$ by a factor of two.  At high
temperatures $c^+_\sigma(q)=2c^-_\sigma(q)$, but as $T$ becomes lower than
$J$, the weight is redistributed so that $c^+_\sigma(q)$ is large at
$|q|<\pi/2$, whereas $c^-_\sigma(q)$ is large at $|q|>\pi/2$, see
Fig.~\ref{fig:c}.  Since the expression (\ref{eq:nu_zeroT}) emphasizes
larger values of $|q|$, the density of states $\nu(-\varepsilon)$ below
the Fermi level increases, whereas $\nu(\varepsilon)$ decreases.

As the temperature is reduced, $c^-_\sigma(\pi)$ grows from 1/6 at
$T\gg J$, Eq.~(\ref{eq:c_incoherent}), to about 0.46 at $T/J\to0$
\cite{private}.  Thus the density of states at negative energies will
increase by nearly a factor of 3.  At the same time the density of
states at positive energies decreases.  Due to the importance of the
subleading contribution (\ref{eq:nu_zeroT_subleading}) the
experimental results cannot be analyzed in terms of the behavior of
$c_\sigma^+(\pi)$, Eq.~(\ref{eq:nu_zeroT_asymptotics}).  However, in
the limit $T\to0$, the density of states at sufficiently low positive
energies will become much smaller than at the negative ones.

To summarize, we predict the density of states $\nu(\varepsilon)$ in a
strongly interacting system of one-dimensional electrons to show
asymmetric behavior illustrated in Fig.~\ref{fig:nu}.  The asymmetry is
strongly temperature-dependent even at $T\ll |\varepsilon|$.  Our results can
be tested in experiments measuring tunneling from a metal tip to the side
of a low-density quantum wire.

\begin{acknowledgments}
  The authors are grateful to K. Penc for helpful discussions.  K.A.M.
  is grateful to RIKEN and the University of Minnesota for their
  hospitality.  This work was supported by the U.S. DOE, Office of
  Science, under Contract No.~\mbox{W-31-109-ENG-38}, by Grant-in-Aid
  for Scientific Research (Grant No.~16GS50219) from MEXT of Japan,
  and by NSF DMR Grants 0237296 and 0439026.
\end{acknowledgments}

\end{document}